# Debye temperature in $YBa_2Cu_3O_x$ as measured from the electron spin–lattice relaxation of doped $Yb^{3+}$ ions

L.K. Aminov [a], V.A. Ivanshin [a,b], I.N. Kurkin [a,*], M.R. Gafurov [a], I.Kh. Salikhov [a], H. Keller [b], M. Gutmann [c]

[a] *MRS Laboratory, Kazan State University, Kremlevskaya Street 18, 420008 Kazan, Russian Federation*
[b] *Physik Institut der Universität Zürich, CH-8057 Zürich, Switzerland*
[c] *Laboratory for Neutron Scattering, ETH Zürich & Paul Scherrer Institut, CH-5232 Villigen, Switzerland*

**Abstract**

The electron spin–lattice relaxation (SLR) times $T_1$ of $Yb^{3+}$ ions were measured from the temperature dependence of electron spin resonance line width in $Y_{0.99}Yb_{0.01}Ba_2Cu_3O_x$ with different oxygen contents. Raman relaxation processes dominate the electron SLR. Derived from the temperature dependence of the SLR rate, the Debye temperature ($\Theta_D$) increases with the critical temperature $T_c$ and oxygen content $x$. This relationship between $T_c$ and $\Theta_D$ can be well understood in terms of the modified Bardeen–Cooper–Schriefer theory of phonon mechanism for a strong electron–phonon coupling.

## 1. Introduction

The Debye characteristic temperature $\Theta_D$ is a quintessential quantity characterizing phonon-related properties of a solid state. This parameter is especially important for the high-temperature superconducting (HTSC) compounds, because the relationship between $\Theta_D$ and the critical temperature ($T_c$) may be very useful for the determination of the possible mechanism of HTSC.

As a rule, the value of $\Theta_D$ is extracted either from elastic constant or specific heat measurements. The experimental data, which were obtained for $\Theta_D$ in $YBa_2Cu_3O_x$ (YBCO) compounds using these methods, are very controversial. So, the value of $\Theta_D$ measured from ultrasonic sound velocities in the work of Ledbetter [1] is much lower than that derived from the specific heat studies [2]. Moreover, as was shown in Ref. [3], $\Theta_D$ in YBCO depends strongly on temperature, and this dependence is usually not observed in other compounds [4]. Therefore, it is very reasonable to estimate value of $\Theta_D$ by other methods.

Here, we present data of $\Theta_D$ obtained from the electron spin–lattice relaxation (SLR) measurements of doped rare-earth $Yb^{3+}$ ions in YBCO ceramic samples with different oxygen contents, $x$, and therefore, with different critical temperatures, $T_c$.

[*] Corresponding author. Tel.: +7-8432-315-506; fax: +7-8432-387-418.
*E-mail address:* igor.kurkin@ksu.ru (I.N. Kurkin).

## 2. Experimental details

The polycrystalline powder $YBa_2Cu_3O_x$ ($x = 6.0, 6.45, 6.67$ and $6.85$) samples were prepared by the standard solid-state reaction technique. Appropriate proportions of $Y_2O_3$, $BaCO_3$, $CuO$ were dried at 400–500°C, mixed and ground thoroughly into a fine powder. The doped $Yb^{3+}$ were added using an oxide $Yb_2O_3$ in a ratio Yb:Y = 1:100. The powder were treated then in a two stage heating process. First, samples with a high oxygen content ($x \simeq 7$) were obtained. These HTSC ceramics were annealed to reduce $x$, which depends strongly on annealing temperature. The oxygen content was defined from the lattice parameter along the crystallographic $c$-axis [5] using X-ray diffraction. The values of $T_c$ for different $x$ (Table 1) were determined from the temperature dependence of microwave absorption in a low magnetic field.

It is very important to investigate a single crystal samples of YBCO, because of their strong anisotropy. In the present work, the YBCO powders were milled (size = 1–3 μm), then mixed with paraffin or epoxy resin and placed in a glass tube in a strong magnetic field ($\geqslant 15$ kG) to prepare a quasi-single-crystal samples (or crystallites). The $c$-axes of these crystallites were predominantly oriented along the direction $C$ of the external magnetic field $H$ after hardening of epoxy resin. ESR spectra were recorded on X-band (~9.25 GHz) IRES-1003 and THN-251 ESR spectrometers in the temperature range from 4 to 120 K.

## 3. Results and discussion

Fig. 1 shows the complicated temperature dependence of the $Yb^{3+}$–ESR peak-to-peak line width $\Delta H_{pp}$ in $YBa_2Cu_3O_{6.85}$ for $H \| C$ with a minimum at the temperature $T_{min}$ and a small step of $\Delta H_{pp}$ near $T_c$. The ESR measurements on this sample were discussed previously at $T < T_{min}$ [6], and in this paper, we report on results above $T_{min}$ only without discussion of the anomaly of $\Delta H_{pp}$ near $T_c$.

We have observed a rapid increase of $\Delta H_{pp}$, as $T$ increases, in all investigated samples for $T > T_{min}$. The dependence of this kind, which is very common for $Yb^{3+}$ ions in single crystals, is caused by a SLR only [7]. We have analyzed the character of the temperature dependence of the SLR rate $T_1^{-1}$ to distinguish the dominating SLR process. Then, the similar approach as explained in Ref. [8] was applied to evaluate the ESR line width caused by SLR only ($\Delta H_{pp}^{SLR}$) from the relation:

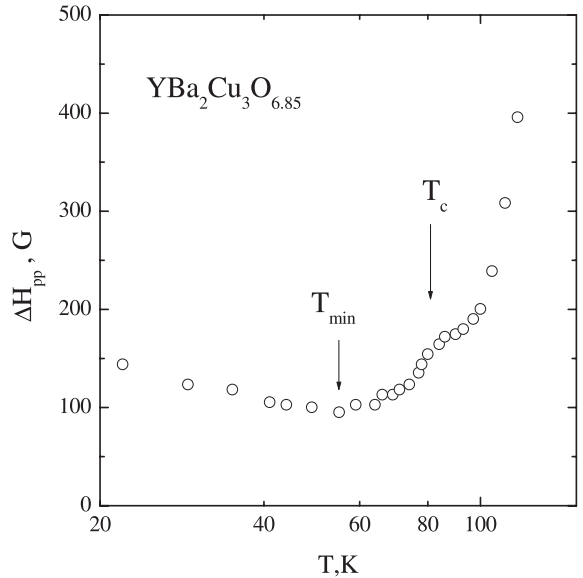

Fig. 1. Temperature dependence of the peak-to-peak ESR line width of $Yb^{3+}$ ions in $YBa_2Cu_3O_{6.85}$, $H \| C$.

Table 1
Critical and Debye temperatures, the minimal ESR line width $\Delta H_{pp}^{min}$, effective $g$-values and the SLR parameter $C$ against oxygen content $x$[a]

| $x$ | 6.85 | 6.67 | 6.45 | 6.0 |
|---|---|---|---|---|
| $T_c$ (K) | 85 | 65 | 40 | – |
| $\Theta_D$ (K) | 450 | 370 | 280 | 250 |
| $\Delta H_{pp}^{min}$ (G) | 95 (55 K) | 90 (45 K) | 75 (25 K) | 95 (20 K) |
| $g$ | 3.0 | 3.1 | 3.15 | 3.45 |
| $C$ ($\times 10^{-7}$) | 1.06 | 4.23 | 15.3 | 50 |

[a] The values of $\Delta H_{pp}^{min}$ were measured at temperatures, which are given in brackets. The $g$-values were estimated at $T \simeq 50$ K with an accuracy of ~20%.

$$(\Delta H_{pp})^2 = \Delta H_{pp}^{SLR} \Delta H_{pp} + (\Delta H_{pp}^{min})^2, \quad (1)$$

where $\Delta H_{pp}^{min}$ is the minimal ESR peak-to-peak line width. The values of $\Delta H_{pp}^{min}$ are given in Table 1 for all investigated samples. The formula (1) is given in another denotations in Ref. [9] and determines an observed line width by the superposition of ESR signals with a Gaussian and Lorentzian line shapes together. We have to take this fact into account in our case, because the ESR line has a Gaussian line shape at $T = T_{min}$ and arises due to inhomogeneities of the crystal electric field (CEF) potential, and the Lorentzian line shape is connected usually to the term $\Delta H_{pp}^{SLR}$.

The SLR rate $T_1^{-1}$ can be expressed from $\Delta H_{pp}^{SLR}$ as follows [10]:

$$T_1^{-1}(s^{-1}) = (\sqrt{3}/2)g\beta\hbar^{-1}\Delta H_{pp}^{SLR}(G)$$
$$\equiv 7.62 \times 10^6 g \Delta H_{pp}^{SLR}(G), \quad (2)$$

where $\beta$ is Bohr's magneton, $\hbar$, the Planck constant, and the g-factor $g = g_\parallel$. The values of $\Delta H_{pp}^{min}$ and $g$, which depend very weakly on temperature, were measured in the temperature range from 25 to 55 K and are given in Table 1. The temperature dependences of the SLR rate $T_1^{-1}$ for the samples with two different oxygen contents $x = 6.85$ and 6.0 are shown in Figs. 2 and 3, respectively. Thus, the $Yb^{3+}$-relaxation rate reads $T_1^{-1} \sim T^n$ ($n = 5$ for $x = 6.85$, $n = 4$ for $x = 6.67$, $n = 3$ for $x = 6.45$ and 6.0). The dependence of such a kind corresponds to the Raman SLR process within the so-called "intermediate" temperature range, if both relations $T \ll \Theta_D$ (then $T_1^{-1} \sim T^9$) and $T \geqslant \Theta_D$ ($T_1^{-1} \sim T^2$) do not take place.

The relaxation rate $T_1^{-1}$ of the Raman SLR process can be written in a common case as follows [7]:

$$T_1^{-1} = CT^9 f(\Theta_D/T), \quad (3)$$

where prefactor $C$ will be discussed later, and the integral function $f(\Theta_D/T) \equiv f(z) \equiv I_8(z)/I_8(\infty)$. The calculated values of the integral $I_8(z) = \int_0^z x^8 e^x/(e^x - 1)^2 dx$ are given [11]. Our experimental data on SLR rate in YBCO can be described quite well using the values of prefactor $C$ and $\Theta_D$ from Table 1.

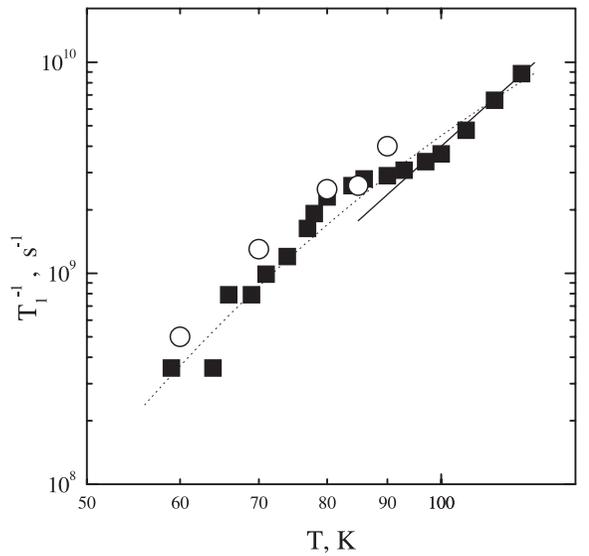

Fig. 2. Temperature dependence of the SLR rate of $Yb^{3+}$ ions in $YBa_2Cu_3O_{6.85}$ measured from the ESR line width (■); empty circles show data of the $^{170}Yb^{3+}$ Mössbauer spectroscopy studies performed by Hodges et al. [13] in $Y_{0.97}Yb_{0.03}Ba_2Cu_3O_7$. Dotted line represents the best fit according to Eq. (3) with $C = 1.06 \times 10^{-7}$ and $\Theta_D = 450$ K. Solid line corresponds to the $T_1^{-1} \sim T^5$ dependence.

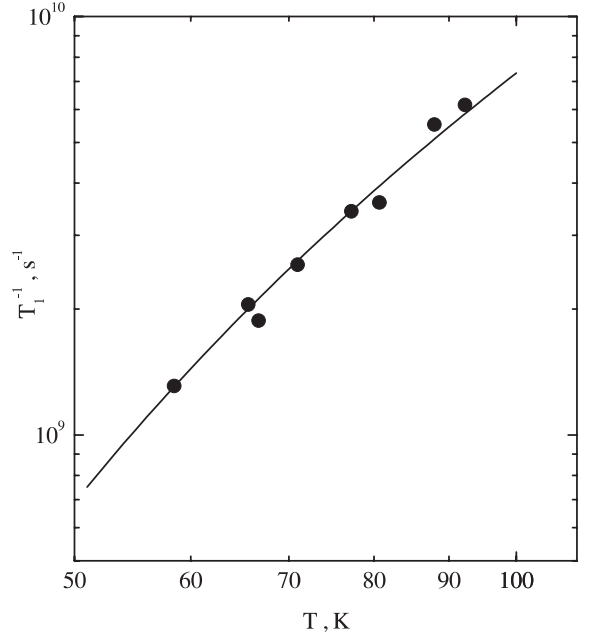

Fig. 3. Temperature dependence of the SLR rate of $Yb^{3+}$ ions in $YBa_2Cu_3O_{6.0}$. The solid curve is the fit to the Eq. (3) with $C = 50 \times 10^{-7}$ and $\Theta_D = 250$ K.

Although the accuracy of determination of $\Theta_D$ was of about ±10%, the estimated value of $\Theta_D = 450$ K for $x = 6.85$ corresponds very well to the literature data [2,3]. The comparison for the samples with $x = 6.67; 6.45$ is not so satisfactory, however, it supports the main conclusion of Refs. [1,12] that $T_c$ is increasing approximately linearly with increase of $\Theta_D$.

In Fig. 2, we show also the temperature dependence of the SLR rate of ytterbium ions extracted from the $^{170}Yb^{3+}$ Mössbauer spectra in a $YBa_2Cu_3O_x$ specimen with $x \approx 7.0$ [13]. A very good correlation to our SLR results seems to confirm that the temperature dependence of the ESR line width at $T > T_{min}$ is caused mainly by the SLR of the $Yb^{3+}$-ions.

The performed studies of the SLR have revealed that by decreasing of oxygen content $x$ and reduction of $T_c$, $\Theta_D$ decreases and the prefactor $C$ on $T^9$ increases. We can evaluate $\Theta_D$ using the well-known expression [4]:

$$\Theta_D = \frac{\hbar}{k}\left(\frac{6\pi^2 q N \rho}{M}\right)^{1/3} v_m, \quad (4)$$

where $v_m$ is the mean sound velocity, $\rho$, the density of material, and $M$, the molecular weight, $q$, the number of atoms in the molecule, and $N$, the Avogadro number. The parameter $C$ can be obtained for the Kramer ions from the relationship [14]:

$$C = \frac{9!\hbar^2}{\pi^3 \rho^2 v_m^{10}}\left(\frac{k}{\hbar}\right)^9 \\ \times \frac{\sum_{nm}|\langle a|V_n^m|c\rangle|^2 \sum_{nm}|\langle c|V_n^m|b\rangle|^2}{\Delta^4} \quad (5)$$

where $\Delta$ is the CEF splitting between the ground and first excited states. The value after sign × is usually approximately equal 1.

Therefore, from relations (4) and (5), it is clear that $\Theta_D$ and $C$ are connected in some way to the sound velocity $v$, and by increasing of $\Theta_D$, $C$ decreases drastically (as $v^{-10}$). This coincides quite well with our experimental data of $\Theta_D$ and $C$ in the samples with $x = 6.85$ and 6.45 (Table 1). Thus, $\Theta_D$ for $x = 6.45$ is of about 1.5 times smaller than one for $x = 6.87$, and the corresponding prefactor $C$ decreases in 15 times in a good accordance with the Eqs. (4) and (5).

Taking into account the formula (4) for $YBa_2Cu_3O_7$ ($\rho = 5.78$ g/cm$^3$, $q = 13$), we have

$$\Theta_D(K) = 1.25 \times 10^{-3} v_m \text{ (cm/s)}. \quad (6)$$

As was shown in Ref. [1], we can find $v_m = 2.89 \times 10^5$ cm/s and $\Theta_D = 362$ K from the expression for mean sound velocity.

$$3v_m^{-3} = v_l^{-3} + 2v_t^{-3}, \quad (7)$$

using the literature data of the YBCO sample with $T_c = 86$ K for the longitudinal and transversal sound velocities ($v_l = 4.25 \times 10^5$ cm/s and $v_t = 2.62 \times 10^5$ cm/s, respectively). Accordingly, we can estimate the value of $C \approx 1.144 \times 10^{-6}$, which correlates well with data of our SLR measurements.

However, the excellent agreement with our SLR experimental results is observed, if we take $v_m = 3.6 \times 10^5$ cm/s. Both calculated values of $\Theta_D = 450$ K and $C \simeq 1.27 \times 10^{-7}$ correspond then accurately to the data of SLR studies.

Finally, a possible mechanism of high-temperature superconductivity can be deduced from the functional dependence of $T_c$ on $\Theta_D$. Thus, Ledbetter [1,12] has applied the usual BCS expression of $T_c$:

$$T_c = 0.86 \Theta_D \exp(-\lambda^{-1}), \quad (8)$$

where an electron–phonon coupling constant $\lambda$ is given by

$$\lambda = C' M \langle \omega^2 \rangle, \quad (9)$$

and has concluded that $\langle \omega^2 \rangle \sim (\Theta_D)^2$ and the BCS formula cannot describe the experimental data [12], where $T_c$ increases with the increase of $\Theta_D$. In fact, combining Eqs. (8) and (9), one can derive that $T_c$ decreases with increase of $\Theta_D$ for a weak electron–phonon coupling ($\lambda < 1$), and $T_c \approx \Theta_D$ in a strong coupling limit $\lambda > 1$. However, both these relations contradict to our and other experiments in YBCO.

The modification of the BCS theory led to the extended McMillan's equation for various classes of strong-coupled superconductors, which is used

extensively for a phonon mechanism of HTSC [15]:

$$T_c = \frac{\Theta_D}{1.45} \exp\left(-\frac{1.04(1+\lambda)}{\lambda - \mu^*(1 + 0.62\lambda)}\right), \quad (10)$$

where $\mu^*$ is parameter of effective Coulomb repulsion. For $\lambda < 1$ and $\mu^* < \lambda$ ( in a weak-coupling limit), Eq. (10) transforms into Eq. (8). The parameters $\lambda$ and $\mu^*$ can be defined for a better description of our measurements for a strong-coupled systems. The corresponding calculations with the formula (10) for $\lambda \leqslant 10$ result in that $T_c \approx 74.7$ K (for $\Theta_D = 450$ K, $\lambda \approx 4$), $T_c \approx 68.0$ K($\Theta_D = 370$ K), and $T_c \approx 56.3$ K ($\Theta_D = 280$ K, $\lambda \approx 10$), if $\lambda = 8 \times 10^5/\Theta_D^2$ and $\mu^* = 0.1$.

In summary, we have presented an alternative possibility to determine Debye temperature in $YBa_2Cu_3O_x$ by means of the electron SLR of doped $Yb^{3+}$-ions, if Raman processes are dominant in SLR. The relationship between critical and Debye temperatures, derived from these studies for different oxygen contents, can be good explained using McMillan's equation in a strong-coupling limit.

## Acknowledgements

The authors acknowledge with thanks the financial support by the Swiss National Science Foundation (Grant no. 7SUPJ048660) and by the Academy of Sciences of Republic Tatarstan. We are very grateful to M.A. Teplov and M.V. Eremin for their continuous interest and discussions of the problem of the electron SLR of paramagnetic centers in HTSC. Our special thanks also go to V.V. Izotov, S.P. Kurzin, and G.V. Mamin for experimental assistance.

## References

[1] H. Ledbetter, J. Mater. Sci. 7 (1992) 2905.
[2] S.E. Inderhees, M.B. Salamon, T.A. Friedmann, D.M. Ginsberg, Phys. Rev. B 36 (1987) 2401.
[3] A. Junod, D. Eckert, T. Graf, E. Kaldis, J. Karpinski, S. Rusiecki, D. Sanchez, G. Triscone, J. Muller, Physica C 168 (1990) 47.
[4] J. Allers, in: U. Mason (Ed.), Physical Acoustics, vol. 3, Part B, Lattice Dynamics, Mir, Moscow, 1968 (Chapter 1).
[5] J.D. Jorgensen, B.W. Veal, A.P. Paulikas, Phys. Rev. B 41 (1990) 1863.
[6] I.N. Kurkin, I.Kh. Salikhov, L.L. Sedov, M.A. Teplov, R.Sh. Zhdanov, JETP 76 (1993) 657.
[7] A.A. Antipin, A.N. Katyshev, I.N. Kurkin, L.Ya. Shekun, Sov. Phys. Sol. State 9 (1967) 3400.
[8] V.A. Ivanshin, M.R. Gafurov, I.N. Kurkin, S.P. Kurzin, A. Shengelaya, H. Keller, M. Gutmann, Physica C 307 (1998) 61.
[9] G.M. Zhidomirov et al., Interpretation of Complicated EPR Spectra, Nauka, Moscow, 1975.
[10] M.D. Kemple, H.J. Stapleton, Phys. Rev. B 5 (1972) 1668.
[11] J.M. Ziman, Proc. Roy. Soc. 226 (1954) 436.
[12] H. Ledbetter, Physica C 235–240 (1994) 1325.
[13] J.A. Hodges, P. Bonville, P. Imbert, G. Jehanno, Physica C 184 (1991) 259.
[14] P.L. Scott, S.D. Jeffries, Phys. Rev. B 127 (1962) 32.
[15] V.L. Ginsburg, D.A. Kirzhnitz (Eds.), Problem of High-Temperature Superconductivity, Nauka, Moscow, 1977.